\documentstyle[preprint,aps]{revtex}

\begin{document}
\draft

\title{ \large Exchange-correlation energy functional constructed from orbital-dependent coupling-constant-averaged pair correlation functions}

\author{Hiroshi Yasuhara$^{1}$, Masahiko Higuchi$^{2}$, and Yoshiyuki Kawazoe$^{1}$}
\address{$^1$Institute for Materials Research, Tohoku University, Katahira, Sendai 980-8577, Japan}
\address{$^2$Department of Physics, Faculty of Science, Shinshu University, Matsumoto 390-8621, Japan}

\date{\today}

\maketitle


\begin{abstract}
An exchange-correlation energy functional 
$ E_{\mathrm xc} $
and the resultant exchange-correlation potential 
$ v_{\mathrm xc}({\bf r}) $
in density-functional theory 
are proposed using orbital-dependent coupling-constant-averaged pair correlation functions,
$ {\bar{g}}^{\sigma \sigma'}( {\bf r, r'} )$
for electronic structure calculations of atoms, molecules, and solids. 
These orbital-dependent 
$ {\bar{g}}^{\sigma \sigma'}( {\bf r, r'} )$
fulfill the symmetric property, 
the Pauli principle and the sum rules. 
In the limit of uniform density 
$ {\bar{g}}^{\sigma \sigma'}( {\bf r, r'} )$
are reduced to the very accurate analogues of the electron liquid 
that are obtained from an interpolation between long- and short-range correlations 
involving the exchange corrections. 
The major contribution of 
$ v_{\mathrm xc}({\bf r}) $
is given in the form of the Coulomb interaction 
with the exchange-Coulomb hole around an electron. 
The present theory not only guarantees local charge neutrality, 
but also reproduces the exact asymptotic form of the exchange potential,
$ v_{\mathrm x}({\bf r}) = - e^2 / r $
for finite systems. 
The present method of dealing with correlations, 
if properly applied to finite systems, 
can give even the asymptotic form of the correlation potential 
$ v_{\mathrm c}({\bf r}) $
of order
$ r^{-4} $
as well as the van der Waals potential of order 
$ r^{-6} $
for large r.
\end{abstract}

\newpage
 
\section{Introduction}


The most important subject in density-functional theory (DFT) [1,2]
is the exploitation of an accurate exchange-correlation energy functional
$ E_{\mathrm xc} $
and the corresponding exchange-correlation potential
$ v_{\mathrm xc}({\bf r}) $
beyond the local-density approximation (LDA) 
and the generalized gradient expansion approximations (GGA's). 
In a previous paper [3] we have proposed 
an orbital-dependent correlation energy functional
$ E_{\mathrm c} \left[ \{ \varphi_i \} \{ \varepsilon_i \} \right] $
which is to be used by the optimized potential method (OPM)[4]
 in combination with the exact orbital-dependent exchange energy functional 
$ E_{\mathrm x} \left[ \{ \varphi_i \} \right] $.
It comprises a direct and exchange pair of second-order perturbation-like terms 
constructed from Kohn-Sham orbitals and Kohn-Sham energies, 
but one of the two Coulomb interactions entering each term 
is replaced by an effective interaction
$ v_{\tiny \textit{eff}}({\bf r}) $
which can in principle contain all the effects of third- and higher-order perturbation terms. 
From the accurate knowledge of long-, intermediate-, and short-range correlations 
of the electron liquid, we have defined such effective interaction
$ v_{\tiny \textit{eff}}({\bf r}) $
for the electron liquid in order to substitute it 
for the orbital-dependent correlation energy functional above.
These orbital-dependent exchange 
and correlation energy functionals depend implicitly on the electron density 
$ n({\bf r}) $
through Kohn-Sham orbitals and Kohn-Sham energies. 
The application of such implicit functionals to DFT 
is founded on the basic assumption that the correlation potential
$ v_{\mathrm c}({\bf r}) $
as well as the exchange potential
$ v_{\mathrm x}({\bf r}) $
will be evaluated by solving an integral equation which relates
$ v_{\mathrm xc}({\bf r}) $
to 
$ \delta E_{\mathrm xc} / \delta \varphi_i ( {\bf r} ) $
and
$ \delta E_{\mathrm xc} / \delta \varepsilon_i $.
The procedure for solving this integral equation has been called 
the optimized potential method (OPM) [4]. 
Actually, the OPM has been applied to the evaluation of the exchange potential
$ v_{\mathrm x}({\bf r}) $
with great success [4,5,6,7].

Very recently, it has, however, been pointed out [8] 
that for finite systems the correlation potential
$ v_{\mathrm c}({\bf r}) $
obtained from a direct application of the OPM to the lowest-order contribution of 
$ E_{\mathrm c} $, i.e., the second-order perturbation terms constructed from Kohn-Sham orbitals 
and Kohn-Sham energies, in fact, becomes divergent in the asymptotic region for large r. 
The unphysical behavior above is caused 
by the presence of unoccupied orbitals and unoccupied energies 
in the perturbation expression for
$ E_{\mathrm c} $. This implies that the OPM, though apparently most promising,
 might not be valid for the evaluation of the correlation potential
$ v_{\mathrm c}({\bf r}) $. 

In this paper we propose an approximation method which enables one to avoid the 
above difficulty in the OPM in order to utilize the effective interaction of the electron 
liquid we have introduced for the evaluation of the correlation potential 
$ v_{\mathrm c}({\bf r}) $
in DFT.
 For this purpose we transform the orbital-dependent exchange 
and the orbital-dependent second-order perturbation-like correlation energy functionals given
 in the previous paper [3] into another expression written 
in terms of the orbital-dependent 
coupling-constant-averaged pair correlation functions, 
$ {\bar{g}}( {\bf r, r'} )$[9,10]. 
This transformation enables one to deal with 
$ E_{\mathrm xc} $
as an explicit functional of the electron density
$ n({\bf r}) $
and the major contribution of 
$ v_{\mathrm xc}({\bf r}) $
is physically appealing since it is given in the form of 
the Coulomb interaction with the exchange-correlation hole 
around an electron at the position 
$ {\bf r} $. 
The remaining terms involving the functional derivative,
$ \delta {\bar{g}}( {\bf r', r''} ) / \delta n(\bf r) $
probably will make a minor contribution to
$ v_{\mathrm xc}({\bf r}) $
since the six-fold integrations over
$ {\bf r'} $
and
$ {\bf r''} $
may be expected to make only a secondary correction to the Hartree potential,
 giving a rather small and nearly uniform modification in the electron density 
$ n({\bf r}) $
of the system. Therefore we deal with these
 terms in the spirit of the local-density approximation (LDA),
 i.e., calculate these terms in the limit of uniform density 
and replace their density dependence by the local density 
$ n({\bf r}) $
of the system. 
It guarantees that the total
$ v_{\mathrm xc}({\bf r}) $
in the present theory is reduced to the LDA in the limit of uniform density.

In section II 
we give an approximate expression for 
the orbital-dependent spin-parallel
 and spin-antiparallel coupling-constant-averaged
 pair correlation functions for use in
 the exchange and correlation functionals in DFT.
 We show that these pair correlation functions not only fulfill 
the inherent symmetric property as well as the requirement 
due to the Pauli principle, but also satisfy the basic sum rules. 
In section III 
we give an approximate 
form of exchange and the correlation potentials 
for practical use. 
The last section is devoted to conclusions and discussions.

\section{  ORBITAL-DEPENDENT COUPLING-CONSTANT-AVERAGED  PAIR CORRELATION FUNCTIONS}   
Let us start with the general expression for
$ E_{\mathrm xc}({\bf r}) $
that Harris and Jones [9] have first introduced in DFT. 
This expression gives a clear physical meaning of exchange and correlation in DFT. 
\begin{equation}
E_{\mathrm xc}  = 
\frac{1}{2} \int \! \! \int d {\bf r} d {\bf r'}
 \frac{e^{2} n({\bf r}) n({\bf r'})}{|{\bf r}-{\bf r'}|}
 ( {\bar{g}}( {\bf r, r'} ) - 1 ) ).
\end{equation}
In Eq.(1), the coupling-constant-averaged pair correlation function 
$ {\bar{g}}( {\bf r, r'} )$
is defined 
using the technique that the electron density
$ n({\bf r}) $
in the real interacting system is maintained 
while the Coulomb interaction among electrons is 
adiabatically switched on as a perturbation. 
The function thus defined,
$ {\bar{g}}( {\bf r, r'} )$
 exhibits general behaviors analogous to 
those of the usual pair correlation function
$ g( {\bf r, r'} )$
as a function of
$ |{\bf r}-{\bf r'}| $. 
The difference of
$ {\bar{g}}( {\bf r, r'} )$
from unity is, however, generally reduced in magnitude compared with that of 
$ g( {\bf r, r'} )$, 
which is noticeable particularly for small separations.  
This is because a lowering in the electron-electron interaction energy 
due to the presence of
$ g( {\bf r, r'} )$
from its value in the Hartree approximation is partially cancelled 
by an increase in the kinetic energy which is caused by correlation and at the same time 
cannot be described in terms of
$ T_{s} $, the kinetic energy of the reference 
non-interacting system with the same
$ n({\bf r}) $
as the real interacting system, 
through the coupling-constant-averaged pair correlation function
$ {\bar{g}}( {\bf r, r'} )$
introduced in DFT. 
From Eq.(1) it is evident that exchange 
and correlation have the effect 
to reduce the contribution of the Hartree energy functional 
from short separations and the effect 
to enhance the contribution from long separations as a counteraction. 
Correlation between electrons not only gives the net reduction 
in the electron-electron interaction energy that 
is gained at the cost of an inevitable increase in the kinetic energy, 
but also plays an important role in the determination of the electron density
$ n({\bf r}) $
of real many-electron systems that are 
for the most part stabilized under the influence of the presence of nuclei. 
Correlation in real systems gives rise to a change in the electron density
$ n({\bf r}) $
in contrast with correlation in the uniform electron liquid.

In addition to the effective interaction
$ v_{\tiny \textit{eff}}({\bf r}) $
we have also calculated with high accuracy the spin-parallel 
and spin-antiparallel coupling-constant-averaged pair correlation functions 
for the electron liquid,
$ {\bar{g}}^{\sigma \sigma}( |{\bf r}-{\bf r'}| )$
and 
$ {\bar{g}}^{\sigma \textit{-}\sigma}( |{\bf r}-{\bf r'}| )$
. We have then ascertained that the difference of
$ {\bar{g}}^{\sigma \sigma}( |{\bf r}-{\bf r'}| )$
from the Hartree-Fock correlation function
$ {g_{\mathrm HF}^{\sigma \sigma}}( |{\bf r}-{\bf r'}| )$
is significantly reduced in comparison with that of
$ {g}^{\sigma \sigma}( |{\bf r}-{\bf r'}| )$
and that the difference of
$ {\bar{g}}^{\sigma \textit{-}\sigma}( |{\bf r}-{\bf r'}| )$
from unity is appreciably reduced in comparison with that of
$ g^{\sigma \sigma}( |{\bf r}-{\bf r'}| )$.
These reductions are due to the correlational increase in the kinetic energy,
which is by definition involved in the coupling-constant-averaged pair correlation functions.
It should, however, be noted 
that these coupling-constant-averaged pair correlation functions
for the electron liquid are inappropriate for use in Eq.(1)
since no orbital-dependent features can be involved in these functions.

Consider the second-order perturbation-like expression 
we have proposed in the previous paper 
for the orbital-dependent correlation energy functional [3]:
\begin{eqnarray}
E_{\mathrm c} = \frac{1}{2} \sum_{i,j}^{occ.} \sum_{a,b}^{unocc.} \left \{ \frac{ \langle \varphi_i \varphi_j | v ( r_{12} ) | \varphi_a \varphi_b  \rangle \langle \varphi_a \varphi_b | v_{\tiny \textit{eff}}( r_{12} ) | \varphi_i \varphi_j  \rangle}{ \varepsilon_i + \varepsilon_j - \varepsilon_a - \varepsilon_b} - \frac{ \langle \varphi_i \varphi_j | v ( r_{12} ) | \varphi_a \varphi_b  \rangle \langle \varphi_a \varphi_b | v_{\tiny \textit{eff}}( r_{12} ) | \varphi_j \varphi_i  \rangle} { \varepsilon_i + \varepsilon_j - \varepsilon_a - \varepsilon_b} \right \}, \nonumber
\end{eqnarray}
%
\begin{eqnarray}
\langle \varphi_i \varphi_j | v ( r_{12} ) | \varphi_a \varphi_b  \rangle = \int \int d {{\bf r}_1} d {{\bf r}_2} {\varphi^*_i}( {{\bf r}_1} ) {\varphi^*_j}( {{\bf r}_2} ) v ( r_{12} ) {\varphi_a}( {{\bf r}_1} ) {\varphi_b}( {{\bf r}_2} ),
\end{eqnarray}
%
\begin{eqnarray}
\langle \varphi_i \varphi_j | v_{\tiny \textit{eff}}( r_{12} ) | \varphi_a \varphi_b  \rangle = \int \int d {{\bf r}_1} d {{\bf r}_2} {\varphi^*_i}( {{\bf r}_1} ) {\varphi^*_j}( {{\bf r}_2} ) v_{\tiny \textit{eff}} ( r_{12} ) {\varphi_a}( {{\bf r}_1} ) {\varphi_b}( {{\bf r}_2} ), \nonumber 
\end{eqnarray}
%
where
$ \varphi_i $,
$ \varphi_a $
and
$ \varepsilon_i $,
$ \varepsilon_a $
are Kohn-Sham orbitals and Kohn-Sham energies, respectively. 
In Eq.(2),
$ v( r_{12} ) = e^2 / |{{\bf r}_{1}}-{{\bf r}_{2}}| $
and the effective interaction
$ v_{\tiny \textit{eff}}({\bf r}) $
is the real-space Fourier transform of the interaction
$ v_{\tiny \textit{eff}}({\bf q}) $
we have defined by an exact expression for the correlation energy of the electron liquid as
\begin{eqnarray}
E_{\mathrm c} = \frac{1}{2} \left( \frac{1}{\Omega} \right)^2 \sum_{\bf q}
\sum_{{\bf p}, {\sigma},{\bf p'}, {\sigma'} } f({\bf p}) f({\bf p'}) v( {\bf q} )
\frac{ ( 1 - f({\bf p + q}) )( 1 - f({\bf p' - q}) ) }
{ \varepsilon_{\bf p} - \varepsilon_{\bf p+q} + \varepsilon_{\bf p'} - \varepsilon_{\bf p'-q}}
 \nonumber\\
\times \left( v_{\tiny \textit{eff}}( {\bf q} ) - \delta_{\sigma \sigma'}
v_{\tiny \textit{eff}}( {\bf -p+p'-q} ) \right),
\end{eqnarray}
where
$ f(\bf p) $
is the Fermi distribution function at zero temperature and
$ \varepsilon_{\bf p} = {\hbar}^{2} {\bf p}^{2} / 2m $;
$ \Omega $ being the volume of the system.
In Eq.(3), long-, intermediate-, and short-range correlations 
arising from all the higher-order perturbation terms 
beyond the second are taken into account in the form of
$ v_{\tiny \textit{eff}}({\bf q}) $. 
We have evaluated the effective interaction
$ v_{\tiny \textit{eff}}({\bf q}) $
by making a sophisticated interpolation 
between long-range correlation in the random-phase approximation (RPA) [11] 
and short-range correlation in the particle-particle ladder approximation [12,13,14] 
such that the corresponding exchange interaction
$ v_{\tiny \textit{eff}}({\bf -p+p'-q} ) $
and its feedback effect on the direct interaction
$ v_{\tiny \textit{eff}}({\bf q}) $
are allowed for in a self-consistent way. 
The new expression for 
$ E_{\mathrm c} $ 
given by Eq.(3) 
not only reproduces available most accurate numerical values [15] 
of the correlation energy within an accuracy of 0.5mRy. per electron 
throughout the entire region of metallic densities, 
but also has the merit of giving spin-antiparallel 
and spin-parallel contributions of the correlation energy separately. 
It is important to realize that the spin-antiparallel contribution occupies 
about 70\% of the total correlation energy 
and the spin-parallel contribution does occupy 
as much as about 30\% throughout the entire region of metallic densities. 
Probably, this ratio will apply to all the valence electrons 
that take part in cohesion of real metals.

It is important to notice that Eq.(2) can be transformed into 
the correlation part of Eq.(1). 
For this purpose we shall start with the spin-dependent version of Eq.(1), 
which enables one to deal with the exchange 
and the correlation energy functionals,
$ E_{\mathrm x} $
and
$ E_{\mathrm c} $, separately. 
\begin{equation}
E_{\mathrm xc}  = 
\frac{1}{2} \sum_{{\sigma}, {\sigma'} } \int \! \! \int d {\bf r} d {\bf r'}
 \frac{e^{2} n_{\sigma}({\bf r}) n_{\sigma'}({\bf r'})}{|{\bf r}-{\bf r'}|}
( {\bar{g}}^{\sigma \sigma'}( {\bf r, r'} ) - 1 ),
\end{equation}
where the spin-parallel coupling-constant-averaged pair correlation function
$ {\bar{g}}^{\sigma \sigma}( {\bf r, r'} ) $
may be split into the Hartree-Fock and the correlational contributions as 
\begin{equation}
{\bar{g}}^{\sigma \sigma}( {\bf r, r'} ) 
= g_{\mathrm HF}^{\sigma \sigma}( {\bf r, r'} )
+ {\bar{g}}_{\mathrm c}^{\sigma \sigma}( {\bf r, r'} ).
\end{equation}
Then the exchange energy functional
$ E_{\mathrm x} $
can be written 
in terms of the Hartree-Fock spin-parallel pair correlation function
$ g_{\mathrm HF}^{\sigma \sigma}( {\bf r, r'} ) $ as
%
\begin{equation}
E_{\mathrm x}  = 
\frac{1}{2} \sum_{\sigma} \int \int d {\bf r} d {\bf r'}
 \frac{e^{2} n_{\sigma}({\bf r}) n_{\sigma}({\bf r'})}{|{\bf r}-{\bf r'}|}
( g_{\mathrm HF}^{\sigma \sigma}( {\bf r, r'} ) - 1 ),
\end{equation}
where  
\begin{equation}
g_{\mathrm HF}^{\sigma \sigma}( {\bf r, r'} ) 
= 1 - \frac{ 1 }{ n_{\sigma}({\bf r}) n_{\sigma}({\bf r'})}
\left| \sum_{i}^{occ.} {\varphi^*_{i \sigma}}( {\bf r} ) 
{\varphi_{i \sigma}}( {\bf r'} ) \right|^2.
\end{equation}
This function is constructed from occupied Kohn-Sham orbitals alone.
On the other hand, the correlation energy functional
$ E_{\mathrm c} $
consists of the spin-antiparallel and spin-parallel contributions as
%
\begin{equation}
E_{\mathrm c} = \sum_{\sigma} \left(
E_{\mathrm c}^{\sigma \textit{-}\sigma} + E_{\mathrm c}^{\sigma \sigma} \right),
\end{equation}
%
\begin{equation}
E_{\mathrm c}^{\sigma \textit{-}\sigma} 
= \frac{1}{2} \int \! \! \int d {\bf r} d {\bf r'}
 \frac{e^{2} n_{\sigma}({\bf r}) n_{\textit{-}\sigma}({\bf r'})}{|{\bf r}-{\bf r'}|}
( {\bar{g}}^{\sigma \textit{-}\sigma}( {\bf r, r'} ) - 1 ),
\end{equation}
%
\begin{equation}
 E_{\mathrm c}^{\sigma \sigma}
= \frac{1}{2} \int \! \! \int d {\bf r} d {\bf r'}
 \frac{e^{2} n_{\sigma}({\bf r}) n_{\sigma}({\bf r'})}{|{\bf r}-{\bf r'}|}
{\bar{g}_{\mathrm c}}^{\sigma \sigma}( {\bf r, r'} ),
\end{equation}
where
$ {\bar{g}}^{\sigma \textit{-}\sigma}( {\bf r, r'} )$
is the spin-antiparallel coupling-constant-averaged pair correlation 
function and
$ {\bar{g}_{\mathrm c}}^{\sigma \sigma}( {\bf r, r'} )$
denotes the correlational part of 
spin-parallel coupling-constant-averaged pair correlation function. 
An inspection of these two contributions of
$ E_{\mathrm c} $
manifests that development of the Coulomb hole 
and further lowering of the Fermi hole 
at short separations both make a contribution to the correlation energy.

Now we can easily transform the second-order perturbation-like expression 
given by Eq.(2) into the above spin-dependent expressions written in terms of
$ {\bar{g}}^{\sigma \textit{-}\sigma}( {\bf r, r'} )$
and
$ {\bar{g}_{\mathrm c}}^{\sigma \sigma}( {\bf r, r'} )$
by paying special attention to a single Coulomb interaction 
entering each of direct and exchange terms in it. 
The two correlation functions,
$ {\bar{g}}^{\sigma \textit{-}\sigma}( {\bf r, r'} )$
and 
$ {\bar{g}_{\mathrm c}}^{\sigma \sigma}( {\bf r, r'} )$
may be identified, respectively, as
%
\begin{eqnarray}
{\bar{g}}^{\sigma \textit{-}\sigma}( {\bf r, r'} ) - 1 =
\frac{ 1 }{ n_{\sigma}({\bf r}) n_{\textit{-}\sigma}({\bf r'})}
\sum_{i,j}^{occ.} \sum_{a,b}^{unocc.}
\frac{ {\varphi^*_{i \sigma}}( {\bf r} ) {\varphi^*_{j \textit{-}\sigma}}( {\bf r'} )
{\varphi_{a \sigma}}( {\bf r} ) {\varphi_{b \textit{-}\sigma}}( {\bf r'} ) }
{ \varepsilon_{i \sigma}+\varepsilon_{j \textit{-}\sigma}
-\varepsilon_{a \sigma}-\varepsilon_{b \textit{-}\sigma} }
\nonumber\\ \times
 \int \! \! \int d {{\bf r}_1} d {{\bf r}_2} 
{\varphi^*_{a \sigma}}( {{\bf r}_1} ) {\varphi^*_{b \textit{-}\sigma}}( {{\bf r}_2} )
v_{\tiny \textit{eff}} ( r_{12} )
{\varphi_{i \sigma}}( {{\bf r}_1} ) {\varphi_{j \textit{-}\sigma}}( {{\bf r}_2} ),
\end{eqnarray}
%
\begin{eqnarray}
{\bar{g}_{\mathrm c}}^{\sigma \sigma}( {\bf r, r'} ) =
\frac{ 1 }{ n_{\sigma}({\bf r}) n_{\sigma}({\bf r'})}
\sum_{i,j}^{occ.} \sum_{a,b}^{unocc.}
\frac{ {\varphi^*_{i \sigma}}( {\bf r} ) {\varphi^*_{j \sigma}}( {\bf r'} )
{\varphi_{a \sigma}}( {\bf r} ) {\varphi_{b \sigma}}( {\bf r'} ) }
{ \varepsilon_{i \sigma}+\varepsilon_{j \sigma}
-\varepsilon_{a \sigma}-\varepsilon_{b \sigma} }
\nonumber\\ \times
 \int \int d {{\bf r}_1} d {{\bf r}_2} 
{\varphi^*_{a \sigma}}( {{\bf r}_1} ) {\varphi^*_{b \sigma}}( {{\bf r}_2} )
v_{\tiny \textit{eff}} ( r_{12} )
\left \{ \varphi_{i \sigma}( {{\bf r}_1} ) \varphi_{j \sigma}( {\bf r}_2 ) - 
\varphi_{j \sigma}( {{\bf r}_1} ) \varphi_{i \sigma}( {\bf r}_2 ) \right \},
\end{eqnarray}
where the electron density with spin $ \sigma $ is given by
\begin{equation}
n_{\sigma}({\bf r}) = \sum_{i}^{occ.} \left | \varphi_{i \sigma}( {\bf r} ) \right|^2.
\end{equation}
Note that these correlation functions are constructed from 
both unoccupied and occupied Kohn-Sham orbitals and Kohn-Sham energies. 
It is important to notice that these correlation functions 
in the limit of uniform density, i.e., if Kohn-Sham orbitals and Kohn-Sham energies 
in Eqs. (11) and (12) are replaced by plane waves and free electron energies, respectively, 
they are reduced to the accurate analogues of 
the electron liquid because of the very definition of the effective interaction
$ v_{\tiny \textit{eff}}({\bf r}) $
that we have introduced.

It is evident from Eqs. (11) and (12) that the present theory 
satisfies the symmetric property inherent in the pair correlation function 
as well as the requirement due to the Pauli principle.
%
\begin{equation}
{\bar{g}}^{\sigma \sigma'}( {\bf r, r'} ) = 
{\bar{g}}^{\sigma' \sigma}( {\bf r', r} ),
\end{equation}
%
\begin{equation}
{\bar{g}}_{\mathrm c}^{\sigma \sigma}( {\bf r, r} ) = 0.
\end{equation}
It is important to realize that the well-known sum rule 
concerning the exchange-Coulomb hole is in fact exhausted 
by the Hartree-Fock exchange hole alone.
%
\begin{equation}
\int d {\bf r'} n_{\sigma} ( {\bf r'} ) \left( 
g_{\mathrm HF}^{\sigma \sigma}( {\bf r, r'} ) - 1
\right) = - 1.
\end{equation}
This is evident from the orthogonality between occupied Kohn-Sham orbitals. 
As a consequence, the Coulomb hole due to correlation 
between electrons with opposite spin and the correlational change 
in the Fermi hole separately have to integrate to zero. 
In other words, correlation due to the Coulomb interaction occurs among electrons, 
for either of two spin orientations, such that local charge neutrality 
is maintained at every position 
$ {\bf r} $
of the system. 
The present theory has the striking merit that the requirements above are fulfilled, 
as is evident from the orthogonality between occupied and unoccupied Kohn-Sham orbitals 
appearing in Eqs.(11) and (12).
\begin{equation}
\int d {\bf r'} n_{\textit{-}\sigma} ( {\bf r'} ) \left( 
{\bar{g}}^{\sigma \textit{-}\sigma}( {\bf r, r'} )
- 1 \right) = 0,
\end{equation}

\begin{equation}
\int d {\bf r'} n_{\sigma} ( {\bf r'} )
{\bar{g}}_{\mathrm c}^{\sigma \sigma}( {\bf r, r'} )
 = 0.
\end{equation}
%
\section{ EXCHANGE AND CORRELATION POTENTIALS }   
In order to obtain the exchange potential 
$ v_{{\mathrm x}\sigma}({\bf r}) $
we evaluate the functional derivative of 
$ E_{\mathrm x} $ 
given by Eq.(6) 
with respect to
$ n_{\sigma}({\bf r}) $. 
The major contribution of 
$ v_{{\mathrm x}\sigma}({\bf r}) $
is given as
%
\begin{equation}
 \int d {\bf r'}
 \frac{e^{2} n_{\sigma}({\bf r'})}{|{\bf r}-{\bf r'}|}
( g_{\mathrm HF}^{\sigma \sigma}( {\bf r, r'} ) - 1 ).
\end{equation}
This is nothing but the Coulomb interaction 
with the bare Fermi hole around an electron located at the position
$ {\bf r} $. 
In the limit of uniform density it is reduced to
$ 3 / 2 {\mu}_{{\mathrm x}\sigma} $
[16] where
$ {\mu}_{{\mathrm x}\sigma} $
denotes the exchange contribution to the chemical potential of the uniform electron liquid. 
The remaining term of
$ v_{{\mathrm x}\sigma}({\bf r}) $
is written as
%
\begin{equation}
\frac{1}{2} \int \int d {\bf r'} d {\bf r''}
 \frac{e^{2} n_{\sigma}({\bf r'}) n_{\sigma}({\bf r''})}{|{\bf r'}-{\bf r''}|}
\frac{ \delta g_{\mathrm HF}^{\sigma \sigma}( {\bf r', r''} ) }
{ \delta n_{\sigma}({\bf r}) }.
\end{equation}
The functional derivative entering Eq.(20) cannot possibly be evaluated 
for arbitrary electronic systems, but Eq.(20) can be analytically evaluated 
in the limit of uniform density. It amounts to be
$ - 1 / 2 {\mu}_{{\mathrm x}\sigma} $[16]. 
It is not likely that Eq.(20) may produce such an important contribution to 
$ v_{{\mathrm x}\sigma}({\bf r}) $
as is comparable to Eq.(19) because of the six-fold integration over
$ {\bf r'} $
and
$ {\bf r''} $. 
Then we approximate this term by
$ - 1 / 2 {\mu}_{{\mathrm x}\sigma}( n_{\sigma}(\bf r) ) $
following the spirit of the LDA.

A similar approximation method may be applied to the correlation potential
$ v_{{\mathrm c}\sigma}({\bf r}) $. 
The potential
$ v_{{\mathrm c}\sigma}({\bf r}) $
in the present theory consists of the major functional derivative of
$ E_{\mathrm c} $
given by Eqs.(8), (9) and (10) with respect to
$ n_{\sigma}({\bf r}) $
and the minor correction as
%
\begin{eqnarray}
v_{{\mathrm c}\sigma}({\bf r}) =
 \int d {\bf r'}
 \frac{e^{2} n_{\textit{-}\sigma}({\bf r'})}{|{\bf r}-{\bf r'}|}
( {\bar{g}}^{\sigma \textit{-}\sigma}( {\bf r, r'} ) - 1 )
+\int d {\bf r'}
\frac{e^{2} n_{\sigma}({\bf r'})}{|{\bf r}-{\bf r'}|}
{\bar{g}_{\mathrm c}}^{\sigma \sigma}( {\bf r, r'} )
+ \Delta {\mu}_{{\mathrm c}\sigma}( n_{\sigma}(\bf r) ).
\end{eqnarray}
The first term is given in the form of the Coulomb interaction 
with the Coulomb hole around an electron located at the position
$ {\bf r} $
and the second term in the form of the Coulomb interaction 
with a correlation-induced change in the Fermi hole around the electron. 
From these expressions it is then evident that the first and second terms 
have the effect to significantly reduce the contribution of the Hartree potential 
from short separations, which is important to the accurate evaluation of the electron density
$ n ( {\bf r} ) $
of the system. 
This is because the screening of each attractive nuclear potential 
by valence electrons is expected to be significantly reduced at short distances 
by the first and second terms in Eq.(21) and the electron density of
$ n ( {\bf r} ) $
may be somewhat enhanced in the immediate vicinity of each nucleus. 
On the other hand, it is not likely that the remaining terms of
$ v_{{\mathrm c}\sigma}({\bf r}) $
may make such an important contribution as is comparable to the first and second terms 
because of the six-fold integrations involved. 
Then we approximate the remaining terms of
$ v_{{\mathrm c}\sigma}({\bf r}) $
in the spirit of the LDA. The last term 
$ \Delta {\mu}_{{\mathrm c}\sigma}( n_{\sigma}({\bf r}) ) $
in Eq.(21) denotes such an approximation to those terms 
which involve the functional derivatives,
$ \delta {\bar{g}}^{\sigma \textit{-}\sigma}( {\bf r', r''} ) / \delta n_{\sigma}(\bf r) $
and
$ \delta {\bar{g}_{\mathrm c}}^{\sigma \sigma}( {\bf r', r''} ) / \delta n_{\sigma}({\bf r}) $.
The coefficient 
$ \Delta $
should be determined such that 
in the limit of uniform dfensity the total correlation potential
$ v_{{\mathrm c}\sigma}({\bf r}) $
is reduced to the correlation contribution to the chemical potential of the electron liquid,
$ {\mu}_{{\mathrm c}\sigma} $.
We think that the contribution of
$ v_{{\mathrm c}\sigma}({\bf r}) $
involving the six-fold integrations will form a background-like potential 
against the Hartree potential and plays a secondary role 
in the occurrence of a correlation-induced change in the electron density
$ n(\bf r) $
of the system. Then we have treated that contribution following 
the spirit of the local-density approximation (LDA).

For the practical application of the present scheme, 
it is a delicate problem how to choose the density parameter in the effective interaction
$ v_{\tiny \textit{eff}}({\bf r}) $
we have borrowed from the electron liquid. 
The best choice of the density parameter will be 
such an optimization as gives the minimum value of the calculated ground state energy. 
Another choice will be the effective average density of valence electrons 
that take part in binding of molecules or in cohesion of solids; 
for metals it will correspond to the nearly uniform density 
realized in the interstitial region outside muffin-tin spheres.

Now all quantities requisite for DFT are available.
The exchange energy functional
$ E_{\mathrm x} $
is exactly given in terms of the Hartree-Fock spin-parallel pair correlation function
$ g_{\mathrm HF}^{\sigma \sigma}( {\bf r, r'} )$
and the correlation energy functional
$ E_{\mathrm c} $
in terms of the spin-antiparallel coupling-constant-averaged pair correlation function
$ {\bar{g}}^{\sigma \textit{-}\sigma}( {\bf r, r'} )$
and the correlation-induced spin-parallel coupling-constant-averaged pair correlation function
$ {\bar{g}}_{\mathrm c}^{\sigma \sigma}( {\bf r, r'} )$. 
Correspondingly, the exchange potential
$ v_{{\mathrm x}\sigma}({\bf r}) $
and the correlation potential
$ v_{{\mathrm c}\sigma}({\bf r}) $
are also given in terms of
$ g_{\mathrm HF}^{\sigma \sigma}( {\bf r, r'} )$, 
$ {\bar{g}}^{\sigma \textit{-}\sigma}( {\bf r, r'} )$,
and 
$ {\bar{g}_{\mathrm c}}^{\sigma \sigma}( {\bf r, r'} )$,
though the minor contribution of
$ v_{{\mathrm x}\sigma}({\bf r}) $
and
$ v_{{\mathrm c}\sigma}({\bf r}) $
is approximately treated in a similar way as in the LDA. 
We have thus avoided the difficulty 
that is encountered in dealing with the correlation potential
$ v_{{\mathrm c}\sigma}({\bf r}) $
according to the OPM by transforming the exchange 
and correlation energy functionals into the expression given by Eq.(4) 
and making some approximations.

In the present scheme, the major contribution of
$ v_{{\mathrm x}\sigma}({\bf r}) $
and 
$ v_{{\mathrm c}\sigma}({\bf r}) $
as well as 
$ E_{\mathrm x} $
and 
$ E_{\mathrm c} $
depends implicitly on Kohn-Sham orbitals and Kohn-Sham energies 
through the coupling-constant-averaged pair correlation functions
$ {\bar{g}}^{\sigma \sigma'}( {\bf r, r'} )$.
The present 
$ v_{{\mathrm c}\sigma}({\bf r}) $,
when applied to finite systems, 
is not divergent for large r, however.
 This is evident from its expression (Eq.(21)) and the sum rules (Eqs. (17) and (18)). 
A much higher degree of self-consistency is required to perform the present scheme than 
in the LDA or GGA's since the coupling-constant-averaged pair correlation functions 
involved are orbital- and energy-dependent.

\section{ CONCLUSIONS AND DISCUSSIONS }

Between metals and non-metals there is an essential difference in long-range correlation, 
or equivalently screening at long distances, 
depending on whether the Fermi surface exists or not. 
We have borrowed the knowledge of long-range correlation 
from the electron liquid in which perfect screening at long distances 
is realized in the presence of the Fermi surface. 
In this sense, the effective potential
$ v_{\tiny \textit{eff}}({\bf r}) $
in the previous paper [3] is properly defined for metals. 
If one performs a similar interpolation between long-range correlation 
in the RPA and short-range correlation 
in the particle-particle ladder approximation for finite systems 
or semiconductors and insulators in order to properly define the effective interaction
$ v_{\tiny \textit{eff}}({\bf r}) $
in the second-order perturbation-like correlation energy functional, 
the resulting effective potential
$ v_{\tiny \textit{eff}}({\bf r}) $
takes the form of 
$ e^2 / {\varepsilon}_{0} r $
at large distances where 
$ {\varepsilon}_{0} $
is a constant of order of the static dielectric constant. 
On the other hand, there is no essential difference in short-range correlation, 
or the analytic form of 
$ v_{\tiny \textit{eff}}({\bf r}) $
at short distances between metals and non-metals. 
The particle-particle ladder approximation [12,13,14] 
gives the adequate description of short-range correlation 
as well as of the cusp condition [17,18,19] on the many-body wavefunction, 
independent of whether the Fermi surface exists or not. 
Therefore the present interpolation method of dealing with correlations 
can give the correct asymptotic form of the correlation potential
$ v_{{\mathrm c}\sigma}({\bf r}) $
of order
$ r^{-4} $
for finite systems as well as the van der Waals potential of order
$ r^{-6} $[20], 
if the effective interaction
$ v_{\tiny \textit{eff}}({\bf r}) $
for finite systems is properly defined as 
$ e^2 / {\varepsilon}_{0} r $
for large r.

From experiences it has been widely observed [10] 
that the LDA has the tendency to underestimate the nucleus-nucleus separation 
in the formation of solids and the tendency to overestimate the binding energy of solids. 
Probably, these shortcomings of the LDA will be, 
for the most part, ascribed to that the screening of each nucleus 
or ion by valence electrons is overestimated at short distances 
and that the ground state energy of the atomic state 
is evaluated too high in comparison with that of the solid state. 
In our view, the adequate inclusion of short-range correlation 
between valence electrons may correct the underestimate 
in the nucleus-nucleus separation in solids. 
It is important to notice that short-range correlation 
between valence electrons has the effect to give less screening of each nucleus 
or ion at short distances and therefore may significantly enhance the electron density
$ n({\bf r}) $
in the immediate vicinity of each nucleus. 
As a counteraction of this enhancement, 
the nucleus-nucleus separation is expected to be somewhat increased.

The present approximation to 
$ E_{\mathrm xc } $
in its construction 
is expected to provide a much more accurate evaluation 
of the ground state energy of both atoms and molecules or solids than the LDA and GGA's. 
In the present
$ E_{\mathrm xc} $,
the self-interaction errors arising from the Hartree energy functional 
are completely cancelled by the Hartree-Fock spin-parallel pair correlation function
$ g_{\mathrm HF}^{\sigma \sigma}( {\bf r, r'} )$, 
though the corresponding exchange potential 
$ v_{{\mathrm x}\sigma}({\bf r}) $
involves a minor approximate part. 
Furthermore, the correlation energy functional
$ E_{\mathrm c} $
in the present scheme cannot be overestimated in magnitude 
since we borrow the knowledge of the effective interaction
$ v_{\tiny \textit{eff}}({\bf r}) $
from the electron liquid instead of the density dependence of the correlation energy. 
The LDA overestimates the correlation energy 
between nearly localized or tightly-binding electrons 
due to the occurrence of the locally high-density region 
caused by these electrons since the correlation energy of the uniform electron liquid 
increases its magnitude like
$ \ln {r_{s}} $
in the high-density region where 
$ {r_{s}} $
is the usual density parameter.

A clear understanding of the difficulty [8] 
that is encountered in the evaluation of
$ v_{{\mathrm c}\sigma}({\bf r}) $
according to the OPM will require a detailed investigation 
of how to make OPM integral equations compatible with Kohn-Sham equations, 
i.e., how to combine the two different self-consistencies in a unified way. 
%
Probably, the unphysical behavior of
$ v_{{\mathrm c}\sigma}({\bf r}) $
reported will indicate 
that the authors may have attempted to combine
the two self-consistencies in a forced way. 
%
%
It will be closely related to the fact that the integral equation for 
$ v_{{\mathrm xc}\sigma}({\bf r}) $, 
if treated in a way consistent with Kohn-Sham equations, 
involves the functional derivative
$ \delta v_{{\mathrm xc}\sigma}({\bf r}) / \delta n_{\sigma'}({\bf r'})$
in it. 
The OPM integral equation in its exact form is equivalent to the Sham-Schluter equation [21]. 
This equation requires that the Kohn-Sham Green's function
$ G^{\mathrm KS}({\bf r,r'},\varepsilon) $
defined for the Kohn-Sham reference non-interacting Hamiltonian 
and the standard many-body theoretical one-electron Green's function
$ G({\bf r,r'},\varepsilon) $
for the same many-electron system give the same electron density
$ n({\bf r}) $.
This requirement can be fulfilled if and only if both
$ G^{\mathrm KS}({\bf r,r'},\varepsilon) $
and 
$ G({\bf r,r'},\varepsilon) $
are exact. 
According to a systematic perturbation expansion of the many-body Hamiltonian 
in which the Kohn-Sham noninteracting reference Hamiltonian 
is dealt with as the unperturbed system, the \textit{n}-th order perturbation energy
$ E_{\mathrm xc}^{(n)} $
for 
$ n \geq 2 $
has turned out to be a functional of not only occupied 
and unoccupied Kohn-Sham orbitals and Kohn-Sham energies, 
but also of
$ v_{{\mathrm xc}\sigma}({\bf r}) $
itself [4]. 
Therefore the \textit{n}-th order integral equation for
$ v_{{\mathrm xc}\sigma}({\bf r}) $
involves the functional derivative
$ \delta v_{{\mathrm xc}\sigma}({\bf r}) / \delta n_{\sigma'}({\bf r'})$
and is extremely difficult to solve [4], 
if one goes beyond the first order perturbation 
$ E_{\mathrm xc}^{(1)} $
equivalent to 
$ E_{\mathrm x} $.

The correlation potential 
$ v_{{\mathrm c}\sigma}({\bf r}) $
is by definition a functional of 
$ n_{\sigma'}({\bf r'}) $.
It is very probable that the accurate spatial dependence of 
$ v_{{\mathrm c}\sigma}({\bf r}) $
cannot be obtained without the detailed knowledge of its functional derivative, 
$ \delta v_{{\mathrm xc}\sigma}({\bf r}) / \delta n_{\sigma'}({\bf r'})$. 
The situation is analogous to the fact that the
$ {\bf p} \textit{-} $
and 
$ {\omega} \textit{-} $
dependence 
of the self-energy operator
$ {\Sigma}_{\sigma}({\bf p},\omega) $
essential to the accurate evaluation of the quasi-particle energy dispersion
$ E({\bf p}) $
for the electron liquid cannot be obtained 
without the detailed knowledge of its functional derivative 
with respect to the one-electron Green's function
$ G_{\sigma'}({\bf p'},\omega') $,  
$ \delta {\Sigma}_{\sigma}({\bf p},\omega) / \delta G_{\sigma'}({\bf p'},\omega') $
[22,23,24]. 
This is because the variation of
$ {\Sigma}_{\sigma}({\bf p},\omega) $
with respect to 
$ {\bf p} $
or 
$ {\omega} $
occurs only through that of each constituent in it,
$ G_{\sigma'}({\bf p'},\omega') $
[25]. Similarly, the accurate 
${\bf r}$-dependence of
$ v_{{\mathrm xc}\sigma}({\bf r}) $
will not be obtained without the detailed knowledge of
$ \delta v_{{\mathrm xc}\sigma}({\bf r}) / \delta n_{\sigma'}({\bf r'}) = {\delta}^2 E_{\mathrm xc} / \delta n_{\sigma}({\bf r}) \delta n_{\sigma'}({\bf r'}) $. 
Probably, the situation in DFT will be much more complicated than in 
the standard many-body formalism. 
This is because the simplicity of Kohn-Sham equations is established in DFT 
at the cost of the extreme complexity of how to obtain
$ E_{\mathrm xc} $
and 
$ v_{{\mathrm xc}\sigma}({\bf r}) $.
From a comparison with the second-order functional derivative of the Hartree energy functional,
$ e^2 / |{\bf r}-{\bf r'}| $
it can be seen that the factor of
$ e^2 ( {\bar{g}}^{\sigma \sigma'}( {\bf r, r'} ) - 1 ) / |{\bf r}-{\bf r'}| $
entering the present
$ E_{\mathrm xc} $
will form a physically appealing and 
important part of the second-order functional derivative,
$ {\delta}^2 E_{\mathrm xc} / \delta n_{\sigma}({\bf r}) \delta n_{\sigma'}({\bf r'})$. 

Finally we would like to stress that the present theory 
gives an adequate description of both the orbital-dependence 
and the non-local dependence of the basic two quantities
$ E_{\mathrm xc} $
and 
$ v_{{\mathrm xc}\sigma}({\bf r}) $
in DFT through the orbital-dependent
$ {\bar{g}}^{\sigma \sigma'}( {\bf r, r'} )$
we have constructed with the help of the knowledge 
of long-, intermediate-, and short-range correlations of the electron liquid 
at metallic densities and that it is a feasible scheme for practical calculations.

%
\vspace{0.5cm}
{\bf ACKNOWLEDGEMENTS}
\vspace{0.5cm}
  
One of the authors (H.Y.) would like to thank Dr. Yasutami Takada 
for kindly informing him of Ref.8 and for several discussions. 
He is grateful to Mr. Kenta Hongo for his sincere help in the preparation of the manuscript. 


\end{document}